# Optically Resolved Exchange Splittings in the Doped Van der Waals Ferromagnet $CrBr_3$:$Yb^{3+}$

Thom J. Snoeren, Kimo Pressler, and Daniel R. Gamelin[*]

*Department of Chemistry, University of Washington, Seattle, WA 98195-1700*

Email: gamelin@uw.edu

**Abstract.** The chromium-trihalides ($CrX_3$; X = Cl, Br, I) have long served as model systems for understanding magnetism and magneto-optics in ionic crystals, and they have recently also emerged as archetypes of magnetic two-dimensional (2D) Van der Waals materials. Although the physical properties of $CrX_3$ compounds have now been explored in great depth, the use of impurity doping to generate new properties remains surprisingly underexplored. Here we report on the magnetic and magneto-optical properties of lanthanide-doped $CrBr_3$, in which $Yb^{3+}$ is introduced as a spin-bearing optical point defect. Narrow-line near-infrared dopant photoluminescence is efficiently sensitized by $CrBr_3$-to-$Yb^{3+}$ energy transfer. Magneto-optical measurements show facile spin manipulation of the paramagnetic $Yb^{3+}$ impurities by the surrounding $CrBr_3$ lattice. Photoluminescence data reveal that the Kramers spin degeneracies of the ground- and excited-state $Yb^{3+}$ doublets are spontaneously lifted by strong magnetic exchange with neighboring $Cr^{3+}$ ions, such that individual $Yb^{3+}$ spin transitions are easily resolved even at zero external magnetic field. The temperature and field dependence of these splittings can be used to probe both local and long-range spin correlations. These results highlight the use of extrinsic optical impurities to add new spin-photonic functionality to this classic 2D Van der Waals magnet.

## I. INTRODUCTION

The intriguing optical and magneto-optical properties of the layered chromium-trihalide ($CrX_3$; X = Cl, Br, I) magnetic insulators have been studied for several decades. As relatively simple ionic crystals, early studies of $CrX_3$ compounds sought to unravel the fundamental origins of their large magnetic rotations of linearly polarized light[1] and the modification of excited energy levels by magnetic states, particularly exchange-related phenomena in the crystal-field transitions of $Cr^{3+}$.[2-4] The recent demonstration of ferromagnetism in monolayer $CrI_3$[5] triggered a rapid resurgence in interest in the chromium trihalides ($CrX_3$).[6-8] Since then, discoveries involving $CrX_3$ compounds have ranged from topological spin textures[9-11] to spin-filtering magnetic tunnel junctions constructed from stacked Van der Waals heterostructures.[12, 13] The broad, featureless *d-d* luminescence and poor spin coherence of $CrX_3$[14-17] do not lend themselves to applications in quantum or classical spin-photonic technologies, however, and they obscure any interplay between magnetic ordering and optical transition energies or electronic structure. Harnessing optical defects in crystals, whether intrinsic or extrinsic, offers a pathway for achieving unique and powerful new functionalities, as demonstrated by the examples of nitrogen-doped diamond, diluted magnetic semiconductors, and the ruby laser.[18-20] In magnetic 2D Van der Waals materials, recent reports have investigated the (magneto-)optical properties of vacancies in CrSBr,[21, 22] but deliberate incorporation of optical impurities remains almost entirely unexplored.

Recently, our group reported on the magneto-optical properties of impurity-doped $CrI_3$.[23] Paramagnetic $Yb^{3+}$ ions added to this lattice are tightly coupled to the surrounding spins and their photoluminescence (PL) allows optical sensing of spontaneous or induced magnetization in bulk $CrI_3$. The specifics of the Yb-Cr coupling remain unclear, however, because anomalously high $Yb^{3+}$ *f*-orbital covalency[17] in $CrI_3$:$Yb^{3+}$ broadens the *f-f* PL, impeding in-depth analysis of the underlying electronic structure.

Here, we report a detailed investigation of the PL of $CrBr_3$ crystals doped with $Yb^{3+}$. $Yb^{3+}$ *f*-orbital covalency in this material is substantially smaller than in $CrI_3$:$Yb^{3+}$.[17] As a consequence, the PL of $CrBr_3$:$Yb^{3+}$ is dramatically narrower and reveals extensive fine structure that has not been resolvable in $CrI_3$:$Yb^{3+}$. Variable-temperature and variable-field measurements allow assignment of the fine structure to specific $Yb^{3+}$ electronic transitions between states whose Kramers spin degeneracies are lifted *via* exchange coupling with the surrounding magnetic $CrBr_3$ lattice. Analysis of these data allows the microscopic $Yb^{3+}$-$Cr^{3+}$ exchange-coupling strength to be quantified and indicates an effective exchange field of ~30 tesla experienced by $Yb^{3+}$ in the low-temperature limit. The observation of well-resolved exchange-split impurity spin levels at zero magnetic field in $CrBr_3$:$Yb^{3+}$ may have interesting ramifications for optical spin preparation and readout, inverse Faraday effect, or other spin-photonic functionalities in this modified magnetic Van der Waals material.

## II. EXPERIMENTAL

*General considerations.* All sample preparation and manipulation was done in a glovebox under an atmosphere of purified dinitrogen.

*Synthesis of $CrBr_3$:$Yb^{3+}$ and $YBr_3$:$Yb^{3+}$ single crystals.* The synthesis of single crystals of $CrBr_3$:$Yb^{3+}$ grown through chemical vapor transport is detailed in Snoeren et al.[17] The synthesis of $YBr_3$:$Yb^{3+}$ was carried out under similar conditions. $YBr_3$ (99.9%, ultra-dry, ThermoFisher Scientific) and Yb metal (99.9%, 40 mesh, BeanTown Chemical) were used as precursors. The precursors were mixed at the desired stoichiometry and sealed in an evacuated quartz ampoule. The ampoule was placed in a tube furnace with the precursors at the center. The sample was kept at 800°C for 4 days.

*Photoluminescence (PL) measurements.* A single crystal of the material was placed between two quartz disks and loaded into a closed-cycle helium cryostat. Measurements were performed under high vacuum ($10^{-6}$ Pa). For $CrBr_3$:$Yb^{3+}$, the sample was excited using a continuous-wave 405 nm (3.06 eV) diode. $YBr_3$:$Yb^{3+}$ was excited using a continuous wave 340 nm (3.64 eV) diode. The diode output was focused down to a spot size of approximately 1 $mm^2$ at an excitation power of approximately 4 $mW/cm^2$. Sample emission was focused into a monochromator with a spectral bandwidth of 0.25 nm for $CrBr_3$:$Yb^{3+}$ and a spectral bandwidth of 0.02 nm for $YBr_3$:$Yb^{3+}$. The 4 K $CrBr_3$:$Yb^{3+}$ PL spectrum shown in the Supplemental Material was collected with a spectral bandwidth of 0.002 nm. Spectra were collected using a $LN_2$-cooled silicon CCD camera and were corrected for instrument response.

*Photoluminescence excitation (PLE) measurements.* PLE measurements were likewise carried out in a closed-cycle helium cryostat under high vacuum. The sample was excited using a broadband tungsten-halogen lamp. The excitation wavelength was selected using a monochromator with a spectral bandwidth of 2.15 nm. Excitation light was focused onto the sample with a spot size of approximately 1 $mm^2$. Sample emission was focused onto a monochromator with a spectral bandwidth of 16 nm and emission counts were monitored using a Hamamatsu InGaAs/InP NIR photomultiplier tube. Spectra were corrected for lamp output.

*Absorption measurements.* Absorption measurements were carried out using a Cary 5000 UV-Vis Spectrophotometer in dual beam mode. Scan rates of 2 nm/s were used with a step size of 1 nm and a spectral bandwidth of 1 nm. Due to the air-sensitive nature of the material, special care was taken to avoid air exposure during the measurement: a single crystal of the material was placed between two quartz disks, spaced with a rubber o-ring to avoid distortion of the crystal while also excluding ambient atmosphere. The ensemble was held together by two metal plates with a circular aperture for light transmission. To ensure all light passed through the sample, a mask with a diameter of about 1 mm was placed on the exterior of one of the quartz disks.

*Magnetic circularly polarized luminescence (MCPL) measurements.* A single crystal was placed between two quartz disks and loaded into a Cryo-Industries SMC-1650 OVT superconducting magneto-optical cryostat oriented in the Faraday configuration. The magnetic field was applied parallel to the $CrBr_3$ easy axis (crystallographic *c* axis). The sample was excited with a 405 nm (3.06 eV) diode at approximately 4 mW/cm$^2$. Sample emission was collected along the magnetic field axis and passed through a liquid-crystal variable retardation plate set at λ/4, followed by a linear polarizer to separate the left- and right-circularly polarized components, and then focused onto a fiber-optic cable. Sample emission was then passed through a monochromator with a spectral bandwidth of 0.25 nm and was collected using a LN$_2$-cooled silicon CCD camera.

Field-sweep measurements were performed by continuously monitoring the sample emission using a monochromator with a spectral bandwidth of 9.9 nm. Emission counts were monitored using a Hamamatsu InGaAs/InP NIR photomultiplier tube with an integration time of 0.1 s. The field was swept at a rate of 1.5 mT/s (0.09 T/min).

Polarization ratios are defined as $\rho = (\sigma^- - \sigma^+)/(\sigma^- + \sigma^+) = (I_L - I_R)/(I_L + I_R)$, following the sign conventions outlined in Piepho and Schatz.[24]

### III. RESULTS AND ANALYSIS

**A. Optical spectroscopy**

Figure 1a shows the room-temperature absorption spectrum of a single-crystal flake of $Yb^{3+}$-doped $CrBr_3$. The absorption spectrum shows two main bands that are assigned to the spin-allowed $Cr^{3+}$ $^4A_{2g} \rightarrow$ $^4T_{2g}$ and $^4T_{1g}$ ligand-field transitions, two shoulders assigned to the spin-forbidden $^4A_{2g} \rightarrow$ $^2T_{1g}$ and $^2T_{2g}$ ligand-field transitions, and a broad band at high energy assigned as a ligand-to-metal charge-transfer (LMCT) transition. All transitions are in good agreement with those previously reported for $CrBr_3$.[4] The weak, parity-forbidden $Yb^{3+}$ *f-f* transitions are not observed due to the low $Yb^{3+}$ doping concentration of ~0.4% relative to the total cation concentration. Figure 1a also shows the 5 K photoluminescence excitation (PLE) spectrum of $CrBr_3$:$Yb^{3+}$, collected by monitoring the $Yb^{3+}$ emission at 1043 nm (*vide infra*). The PLE spectrum shows the same features as the absorption spectrum, with narrower peaks due to the lower temperature, demonstrating efficient sensitization of $Yb^{3+}$ PL by $CrBr_3$.

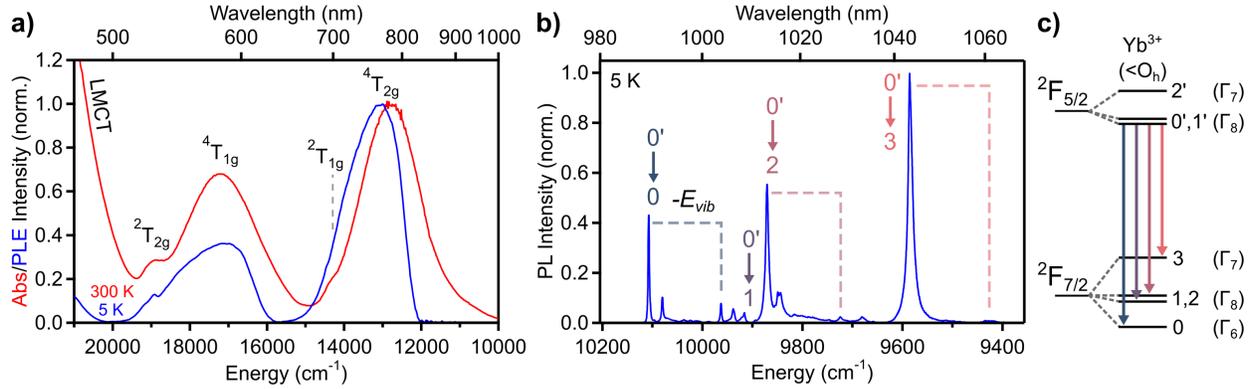

**Figure 1.** **(a)** 300 K absorption spectrum (red) of a $CrBr_3$:$Yb^{3+}$ (0.4%) single crystal. Several $Cr^{3+}$ $^4A_{2g} \rightarrow$ $^{2S+1}\Gamma$ transitions are labeled. The same transitions are observed in the 5 K $Yb^{3+}$ PLE spectrum ($\lambda_{em}$ = 1043 nm, see panel (b)), indicating energy transfer between $Cr^{3+}$ and $Yb^{3+}$. The shifts in PLE peak positions relative to the absorption measurement are due to the measurement temperature. **(b)** 5 K PL spectrum ($\lambda_{ex}$ = 405 nm) of $CrBr_3$:$Yb^{3+}$ (0.4%), showing sharp near-IR $Yb^{3+}$ emission peaks. The electronic origins are labeled and vibronic features are indicated with dashed lines. **(c)** Schematic overview of the observed transitions in the $Yb^{3+}$ PL spectrum. A slightly distorted octahedral crystal field splits the orbital degeneracies of the $\Gamma_8$ levels.

Figure 1b shows the sensitized $Yb^{3+}$ PL spectrum measured at 5 K. The spectrum is characterized by very narrow linewidths – the highest energy peak has a full width at half maximum (FWHM) of ~1.1 cm$^{-1}$ (0.14 meV, see Figure S1[25]). This is in contrast to the PL spectrum of $CrI_3$:$Yb^{3+}$, where the FWHM of the narrowest feature is about 30x larger, even at liquid helium temperatures.[23] Aside from the assigned electronic origins, separate sets of peaks (indicated by dashed lines) are found at fixed energy spacings of ~145 cm$^{-1}$ from the origins. This spacing does not exactly match an $A_{1g}$ vibration in either $CrBr_3$[26] or $YbBr_3$;[27] these peaks are assigned to coupling of the electronic transitions with local $[YbBr_6]^{3-}$ cluster vibrations. The main $Yb^{3+}$ transitions are schematically indicated in Figure 1c, where the orbital degeneracies of the $\Gamma_8$ spin-orbit states in both the $^2F_{7/2}$ and $^2F_{5/2}$ multiplets are lifted as a result of a small reduction in site symmetry from $O_h$.

**B. Magneto-optical properties**

The top part of Figure 2a shows low-temperature magnetic circularly polarized luminescence (MCPL) spectra of $CrBr_3$:$Yb^{3+}$ (0.4%) measured for $\sigma^-$ and $\sigma^+$ circular polarizations at an external magnetic field of 6.0 T. At this field, the lattice is fully magnetized and the polarization ratio, defined as $\rho = (\sigma^- - \sigma^+)/(\sigma^- + \sigma^+)$,[24] is maximized. The bottom half of Figure 2a plots the temperature dependence of $\rho$ for the full spectrum measured at 0.5 T. It is evident that $\rho$ decreases in magnitude as the temperature increases, and almost vanishes at temperatures above $T_C$. Simultaneously, most peaks show a blueshift in the center of gravity, in part due to the appearance of hot bands.

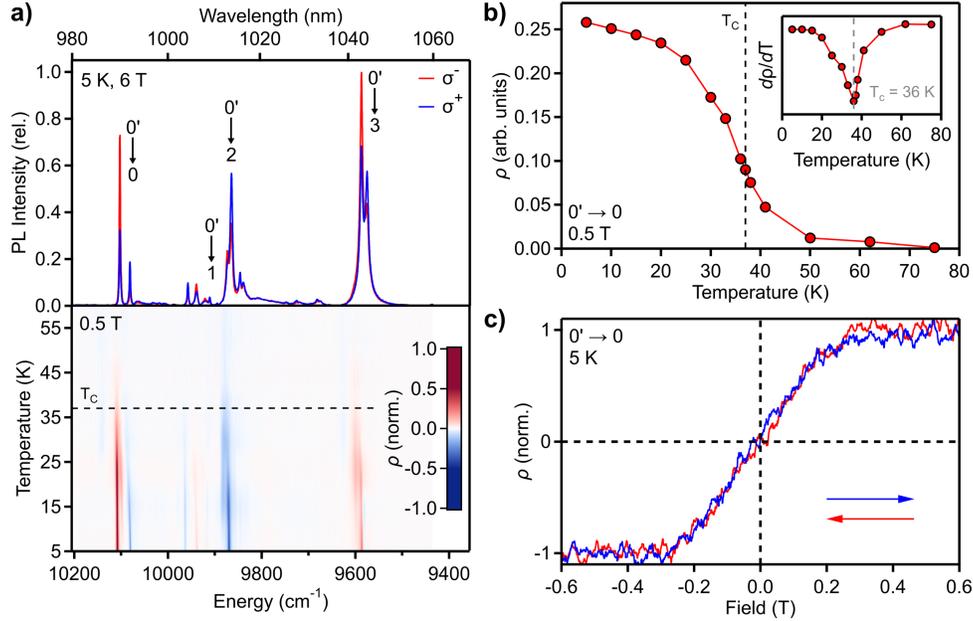

**Figure 2. (a)** Top: MCPL spectra of $CrBr_3$:$Yb^{3+}$ (0.4%) showing left- and right-circularly polarized emission, denoted $\sigma^-$ and $\sigma^+$, respectively. Spectra were collected at $B_{ext}$ = 6.0 T. Bottom: temperature dependence of the polarization ratio ($\rho$) of the full $Yb^{3+}$ PL spectrum collected at $B_{ext}$ = 0.5 T, where $\rho = (\sigma^- - \sigma^+)/(\sigma^- + \sigma^+)$. **(b)** Polarization ratio of the highest-energy $Yb^{3+}$ 0' → 0 PL peak ($A_1$, see section C) plotted as a function of temperature. The derivative of these data shows a minimum at 36 K, in close agreement with the reported $CrBr_3$ Curie temperature of 37 K. **(c)** Normalized polarization ratio of the $Yb^{3+}$ 0' → 0 PL plotted as a function of external magnetic field, $B_{ext}$. The saturation field of ~0.3 T agrees well with reported $CrBr_3$ magnetic data.

Figure 2b plots the polarization ratio of the highest-energy transition (~10110 cm$^{-1}$) as a function of temperature, highlighting the rapid drop in $\rho$ as the temperature approaches the $CrBr_3$ Curie temperature. Some polarization is observed above $T_C$, up to a temperature of around 50 K. The inset shows the derivative of $\rho$ with respect to temperature, which has a minimum at 36 K, in good agreement with the reported $CrBr_3$ Curie temperature of 37 K.[28]

Figure 2c shows a magnetic field-sweep hysteresis loop obtained by monitoring the polarization ratio of the 10110 cm$^{-1}$ Yb$^{3+}$ emission as a function of external magnetic field. The saturation field of ~0.3 T agrees well with magnetization measurements of bulk CrBr$_3$.[29, 30] As previously shown for CrI$_3$:Yb$^{3+}$,[23] this correlation of Yb$^{3+}$ and Cr$^{3+}$ magnetization indicates pinning of the Yb$^{3+}$ spins to the magnetic host lattice under these conditions. Due to the similarity in magneto-optical properties of CrBr$_3$:Yb$^{3+}$ and CrI$_3$:Yb$^{3+}$, we can make use of the highly resolved spectral data on CrBr$_3$:Yb$^{3+}$ to deepen our understanding of both materials.

## C. Yb$^{3+}$ exchange splitting

Figure 3a shows an enlarged view of the Yb$^{3+}$ 0' → 0 transition, plotting spectra in this region collected at various temperatures up to slightly above $T_C$. Full spectra are shown in Figures S2 and S3[25]. At 5 K, only peaks A$_1$ and A$_2$ are clearly visible, separated by 27 cm$^{-1}$. At 10 K, two new peaks (B$_1$ and B$_2$) appear, and these are also separated from one another by 27 cm$^{-1}$. The energy spacing between A$_1$ and B$_1$ is equal to the energy spacing between A$_2$ and B$_2$, 16 cm$^{-1}$ at 5 K. At 15 K, A$_2$ and B$_2$ begin to shift to higher energies and the linewidths of all transitions increase. Above 25 K, another peak (C$_1$) starts to grow at ~32 cm$^{-1}$ above A$_1$. Following the splitting patterns of the other peaks described above, a sixth peak (C$_2$) may also be expected 27 cm$^{-1}$ below C$_1$, but this peak is not clearly observed and may be present only as a shoulder on A$_1$. Above 30 K, all transitions become severely broadened and individual features are poorly resolved.

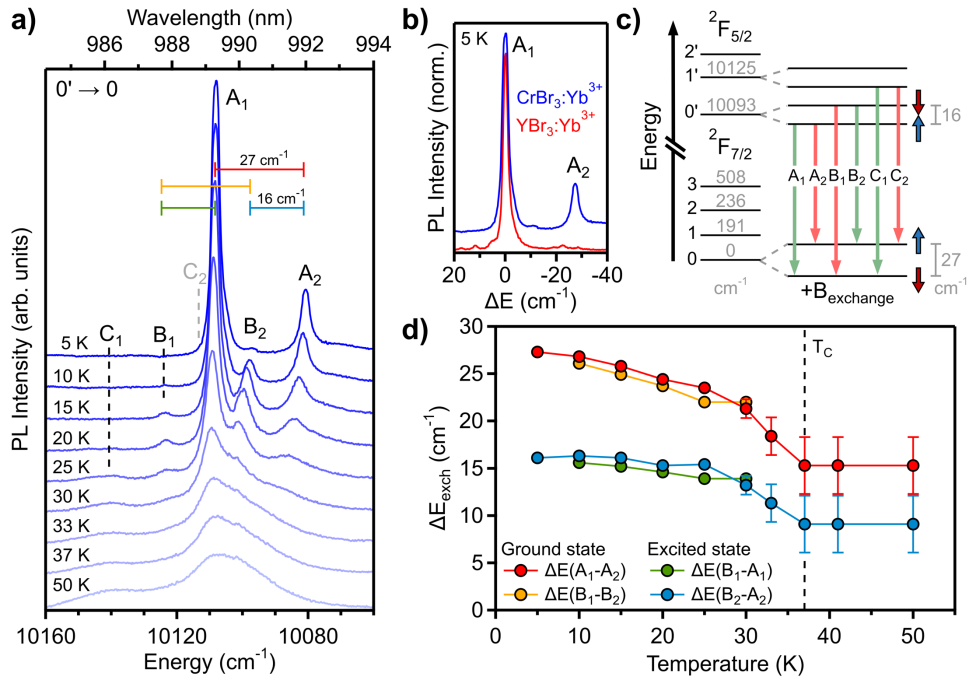

**Figure 3. (a)** Variable-temperature PL spectra of $CrBr_3$:$Yb^{3+}$ measured in the region of the 0' → 0 transition. Fine-structure peaks and relevant splitting energies are labeled. The peaks shift as the temperature is increased, and the fine structure largely becomes unresolved when $T$ approaches $T_C$. Colored bars indicate the energy spacings tracked in panel (d). **(b)** Enlarged view of the 0' → 0 transition in $CrBr_3$:$Yb^{3+}$ (blue) and a reference sample involving a non-magnetic host lattice, $YBr_3$:$Yb^{3+}$ (red), measured at 5 K. The spectra are normalized and offset for comparison. The $A_2$ peak, which results from exchange splitting of the ground state (0), is absent in $YBr_3$:$Yb^{3+}$. **(c)** Energy diagram for $Yb^{3+}$ showing the 5 K exchange splittings of ground- and excited-state Kramers doublets at zero field, based on the observations of panel (a). The corresponding PL transitions of panel (a) are indicated. The colors indicate if the transition is allowed (green) or forbidden (red), as judged from peak intensities. Individual $Yb^{3+}$ spin states are indicated, as explained in section D. **(d)** Ground-state (0) and first-excited-state (0') exchange splittings ($\Delta E_{exch}$) at zero field plotted as a function of temperature, as determined from the PL peak energies in panel (a). The decrease in $\Delta E_{exch}$ with increasing temperature is attributed to loss of lattice spin correlation at higher temperatures.

The rich fine structure in the $CrBr_3$:$Yb^{3+}$ PL spectrum can be explained by the presence of magnetic exchange splittings in both the ground and excited crystal-field states of $Yb^{3+}$ at zero external field. To illustrate the magnetic character of these peaks, crystals of $YBr_3$:$Yb^{3+}$ were prepared and their 5 K PL was investigated. $YBr_3$ is isostructural to $CrBr_3$ (space group $R\bar{3}h$), but due to the full $d$ shell of $Y^{3+}$, the lattice is diamagnetic. The two spectra show the same main features, with the $YBr_3$:$Yb^{3+}$ spectrum shifted to slightly higher energies and spaced closer together due to a reduced crystal field compared to $CrBr_3$:$Yb^{3+}$ (see Figure S5[25]). Both observations can be attributed to the longer metal-ligand bond lengths of $YBr_3$:$Yb^{3+}$. Figure 3b compares the 5 K PL spectra of $CrBr_3$:$Yb^{3+}$ and $YBr_3$:$Yb^{3+}$ zoomed in at the first $Yb^{3+}$ electronic origin (plotted as energy shift from the respective 0' → 0 origin), highlighting the fact that the $YBr_3$:$Yb^{3+}$ spectrum shows no feature analogous to the $A_2$ peak of $CrBr_3$:$Yb^{3+}$. Similar differences are identified for each of the electronic origins when comparing the full spectra (Figure S5[25]), and higher-temperature spectra show the absence of hot bands analogous to the B peaks in $YBr_3$:$Yb^{3+}$ (Figure S6[25]). These results support the assertion that the rich structure described in Figure 3a arises from magnetic exchange. Notably, the low-temperature linewidth of $CrBr_3$:$Yb^{3+}$ in Figure 3b is essentially indistinguishable from that of $Yb^{3+}$ in the non-magnetic $YBr_3$ lattice, consistent with minimal spin disorder in $CrBr_3$:$Yb^{3+}$ at 5 K. At higher temperatures, spin fluctuations cause the $CrBr_3$:$Yb^{3+}$ linewidths to increase rapidly until $T_C$ is reached, while the linewidths of the diamagnetic $YBr_3$:$Yb^{3+}$ increase only marginally over the same temperature range (Figure S6[25]). This result highlights the sensitivity of $Yb^{3+}$ optical transitions to spin fluctuations in $CrBr_3$.

Figure 3c summarizes the results of Figure 3a, assigning each peak to a specific transition between exchange-split $Yb^{3+}$ levels. From this diagram, it is clear that the splitting between peaks $A_1$ and $A_2$ (or $B_1$ and $B_2$) reveals the ground-state exchange splitting ($\Delta E_{exch}$), whereas the splitting between peaks $A_1$ and $B_1$ (or $A_2$ and $B_2$) corresponds to $\Delta E_{exch}$ of the 0' excited state. Figure 3d plots the ground-state (0) and first-excited-state (0') splittings ($\Delta E_{exch}$) at zero external magnetic field as a function of temperature. For both states, $\Delta E_{exch}$ decreases with increasing temperature, consistent with a decrease in magnetization of the host lattice at higher temperatures. By 37 K, the splitting is unresolved and only a broad shoulder is observed. The spectrum at 50 K (above $T_C$) is nearly indistinguishable from that at 37 K, indicating a discontinuity at $T_C$ and confirming the magnetic source of these splittings. For both the ground and excited states, $\Delta E_{exch}$ above $T_C$ is around half as large as at 4 K.

### D. $Yb^{3+}$ spin alignment in $CrBr_3$

Figure 4a shows $CrBr_3$:$Yb^{3+}$ 0' → 0 PL spectra collected at different external magnetic fields, measured at 18 K where $B_1$ and $B_2$ peaks are visible. Peaks $B_1$ and $A_2$ do not shift with field, whereas peaks $A_1$ and $B_2$ shift in opposite directions. The absence of a conventional Zeeman splitting of any of the individual peaks – which would amount to ~1 $cm^{-1}$/T assuming a *g* value of ~2.0 – is consistent with the spin-degeneracy of the $Yb^{3+}$ sublevels having already been lifted at zero external field. The remaining regions of the spectrum match this interpretation (see Figure S7[25]).

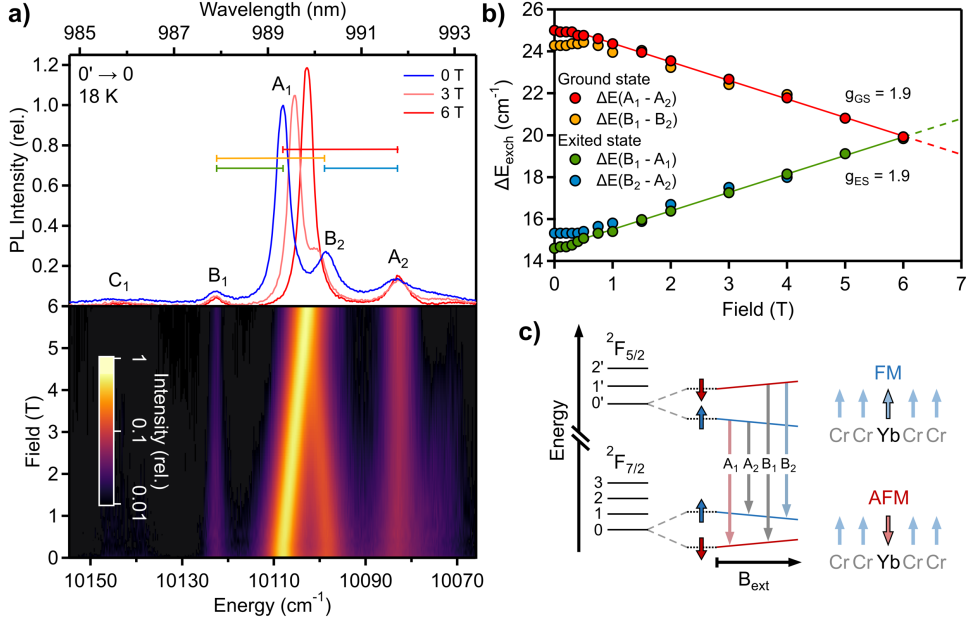

**Figure 4. (a)** Top: PL spectra of $CrBr_3$:$Yb^{3+}$ (0.4%) measured in the region of the 0' → 0 transition at 18 K and at various external magnetic fields ($B_{ext} \parallel c$). Colored bars indicate the energy spacings tracked in panel (b). Bottom: False-color plot of PL in the same region vs $B_{ext}$. **(b)** Splittings of the $Yb^{3+}$ Kramers doublets ($\Delta E_{exch}$) plotted as a function of $B_{ext}$. The solid lines represent linear fits of the data above 0.5 T. **(c)** Energy diagram summarizing the dependence of the $Yb^{3+}$ states on $B_{ext}$, with $Yb^{3+}$ spins coupled antiferromagnetically to the surrounding $Cr^{3+}$ lattice in the ground state and ferromagnetically in the excited state. The colors of the arrows indicate whether the transition redshifts, blueshifts, or is insensitive to the magnetic field. The splittings at $B_{ext}$ = 0 are the exchange splittings ($\Delta E_{exch}$) summarized in Figure 3.

Using the peak assignments of Figure 3c, Figure 4b plots the ground- and excited-state splitting energies ($\Delta E_{exch}$) as a function of external magnetic field. Up to ~0.3 T, the $Yb^{3+}$ splittings are nearly insensitive to the external magnetic field, attributed to the presence of cancelling magnetic domains at zero field: the external field's effect on $Yb^{3+}$ spins in domains magnetized parallel to that field is cancelled out by its effect on $Yb^{3+}$ spins in domains having the opposite magnetization. At field strengths greater than ~0.3 T, the lattice is fully magnetized (Figure 2c) and $\Delta E_{exch}$ of the $Yb^{3+}$ ground state decreases while $\Delta E_{exch}$ of the excited state increases, with slopes corresponding to $g$ values of similar magnitude (~1.9, eq 1) but opposite signs. The field-dependence of the $Yb^{3+}$ energies is summarized schematically in Figure 4c. From $\Delta E_{exch}$ and the $g$ value obtained from Figure 4b, the exchange field experienced by $Yb^{3+}$ can be estimated to be ~30 T at 5 K.

$$g = \left| \frac{1}{\mu_B} \frac{\partial \Delta E_{exch}}{\partial B} \right| \qquad (1)$$

The difference in sign between ground- and excited-state $g$ values indicates that the $Yb^{3+}$ spins reverse their orientation in the magnetic field between the ground and emissive excited states. This change in $Yb^{3+}$ spin orientation is a direct consequence of the difference in total angular momentum $J$ between the $^2F_{7/2}$ and $^2F_{5/2}$ parent terms.[31] Importantly, the magnetic field dependence of the individual spin sublevels also reveals that the $Yb^{3+}$ spins are antiferromagnetically coupled with $CrBr_3$ in the $Yb^{3+}$ ground state but are ferromagnetically coupled with $CrBr_3$ in the emissive $Yb^{3+}$ excited state (Figure 4c). Note that this analysis does not provide information on any possible canting of the $Yb^{3+}$ spins relative to $Cr^{3+}$ spins or to the $CrBr_3$ plane, and further experiments would be required to determine the full details of the $Yb^{3+}$ spin tensor in this material.

The low-temperature value for the ground-state exchange splitting can be used to estimate the Yb-Cr pairwise exchange coupling parameter $J_{Yb-Cr}$, using the exchange Hamiltonian $H = -\sum J_{ij} S_i \cdot S_j$. In the absence of an external field, $H_{eff} = H_{mol}$, so we can rewrite the Curie-Weiss molecular field expression for $H_{eff}$[32] to get eq 2.

$$J_{Yb-Cr} = \frac{g\mu_B H_{mol}^{Yb}}{z\langle S_{Cr}\rangle} = \frac{\Delta E_{exch}}{z\langle S_{Cr}\rangle} \quad (2)$$

In this equation, $z$ is the number of nearest-neighbor $Cr^{3+}$ ions (3) surrounding a given $Yb^{3+}$ impurity and $\langle S_{Cr}\rangle$ is the $Cr^{3+}$ spin expectation value, which has an absolute value of 3/2 in the fully ordered state of $CrBr_3$ (low-temperature limit). Using an extrapolated 0 K ground-state exchange splitting of 27.6 cm$^{-1}$ (Figure 3d) and accounting for the antiferromagnetic alignment of $Yb^{3+}$, $J_{Yb-Cr}$ is calculated to equal -6.2 cm$^{-1}$ (-0.77 meV). This value is half as large as the experimental nearest-neighbor $J_{Cr-Cr}$ found for $CrBr_3$ (+12 cm$^{-1}$ (+1.5 meV)).[33, 34] Lanthanide $f$-orbital exchange interactions are typically weak due to high shielding, but this comparison suggests considerable overlap between $Yb^{3+}$ $4f$ and $Cr^{3+}$ $3d$ wavefunctions in $CrBr_3:Yb^{3+}$.

The data in Figure 3d show that for both the 0 and 0' states of $Yb^{3+}$, $\Delta E_{exch}$ at high temperature (paramagnetic) equals ~50% of its value at the lowest temperature (ferromagnetic). The persistence of a sizeable exchange splitting even above $T_C$ can be understood by considering that the Yb-Cr interaction is short-range, depending only on the immediately neighboring $Cr^{3+}$ ions, and it is not affected by the presence or absence of long-range spin correlation. Thus, although $\langle S_{Cr}\rangle = 0$ in the high-temperature (paramagnetic) regime at zero field, the three nearest-neighbor $Cr^{3+}$ ions exchange coupled to a given $Yb^{3+}$ dopant may have non-zero net spin. A 2D Ising spin lattice has two phases: ferromagnetic (*e.g.*, all $Cr^{3+}$ spins have $S_z = +3/2$) and paramagnetic (spins disordered, *e.g.*, $S_z$ is randomly +3/2 or −3/2 for $Cr^{3+}$). Considering all possible spin configurations of the cluster of three neighboring $Cr^{3+}$ ions acting independently in the Ising paramagnetic phase, two have $S_{3Cr} = 9/2$ and six have $S_{3Cr} = 3/2$. Due to interconversion among these configurations, the observed $Yb^{3+}$ $\Delta E_{exch}$ will reflect the average magnitude of $S_{3Cr}$. Thus, one predicts a ratio of $\Delta E_{exch}(PM)/\Delta E_{exch}(FM) = 1/2$ for $CrBr_3:Yb^{3+}$, agreeing well with our experimental observations. This same consideration holds even in more complicated (and realistic) descriptions of the magnetism of this extended lattice, but the fact that the value of $\Delta E_{exch}$ above $T_C$ predicted in this simple Ising approximation is similar to the one observed experimentally (Figure 3d) suggests that this system remains strongly anisotropic up to at least ~50 K. The exchange splittings observed above $T_C$ in Figure 3d therefore reflect non-cancelling short-range coupling between $Yb^{3+}$ and its three nearest-neighbor $Cr^{3+}$ ions, whereas the PL line broadening in this regime reflects spin disorder among these nearest-neighbor $Cr^{3+}$ ions.

## IV. CONCLUSION

We have shown that incorporation of $Yb^{3+}$ impurities into $CrBr_3$ yields sharp $Yb^{3+}$ *f-f* emission centered at ~10,000 cm$^{-1}$, sensitized through $Cr^{3+}$ *d-d* absorption. Variable-field MCPL measurements show that $Yb^{3+}$ spins are pinned to the $CrBr_3$ magnetization, providing unusually facile control over the dopant's magnetization. The highly resolved *f-f* PL spectra allow direct optical quantification of $Yb^{3+}$-$Cr^{3+}$ exchange splittings, revealing them to vastly exceed $kT$ at 5 K and showing that $Yb^{3+}$-$Cr^{3+}$ coupling inverts from antiferromagnetic to ferromagnetic upon photoexcitation. The large spontaneous exchange splittings observed in $CrBr_3:Yb^{3+}$ present unique opportunities for optical spin manipulation of point defects in the absence of external magnetic fields, or for *in situ* monitoring of short-range spin fluctuations and magnetic phase transitions. Such optically addressable spin defects may also have interesting ramifications for influencing spin textures and domain wall pinning in $CrBr_3$ itself, for generating localized strain effects, or for adding spin-photonic functionality to exfoliated or stacked Van der Waals nanostructures.

## ACKNOWLEDGMENTS

This work was primarily supported by the University of Washington Molecular Engineering

Materials Center, a U.S. National Science Foundation Materials Research Science and Engineering Center (DMR-2308979). Additional support was received from the UW Clean Energy Institute (graduate fellowship to T.J.S.). The authors thank Prof. Rémi Beaulac for valuable discussions. Part of this work was conducted at the Molecular Analysis Facility, a National Nanotechnology Coordinated Infrastructure (NNCI) site at the University of Washington that is supported in part by the National Science Foundation (NNCI-1542101 and NNCI-2025489), the University of Washington, the Molecular Engineering & Sciences Institute, the Clean Energy Institute, and the National Institutes of Health.

# Optically Resolved Exchange Splittings in the Doped Van der Waals Ferromagnet CrBr$_3$:Yb$^{3+}$

*Thom J. Snoeren, Kimo Pressler, and Daniel R. Gamelin*[*]
*Department of Chemistry, University of Washington, Seattle, WA 98195-1700*
Email: gamelin@uw.edu

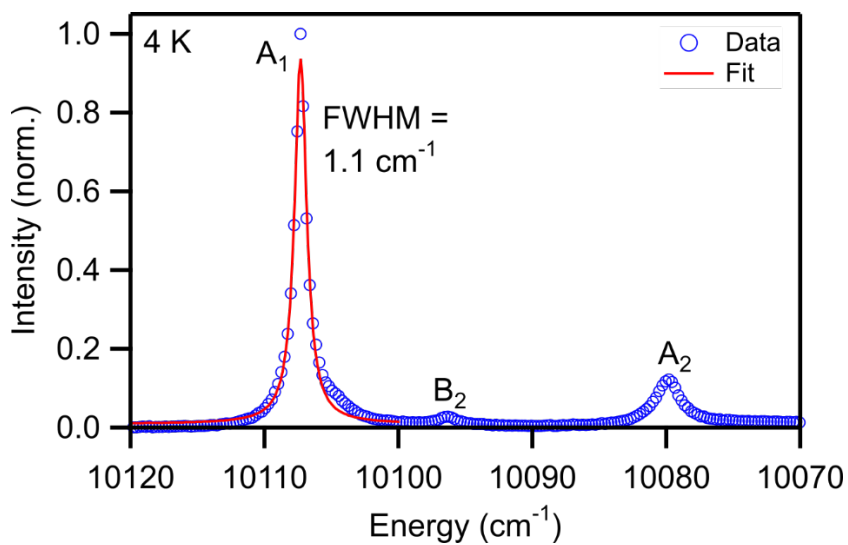

**Figure S1.** Zoom-in on the 0' → 0 transition in the 4 K PL spectrum of CrBr$_3$:Yb$^{3+}$ collected at a spectral bandwidth of 0.002 nm. A Lorentzian fit of peak A$_1$ (red) gives a full width at half max (FWHM) of 1.1 cm$^{-1}$ (0.14 meV).

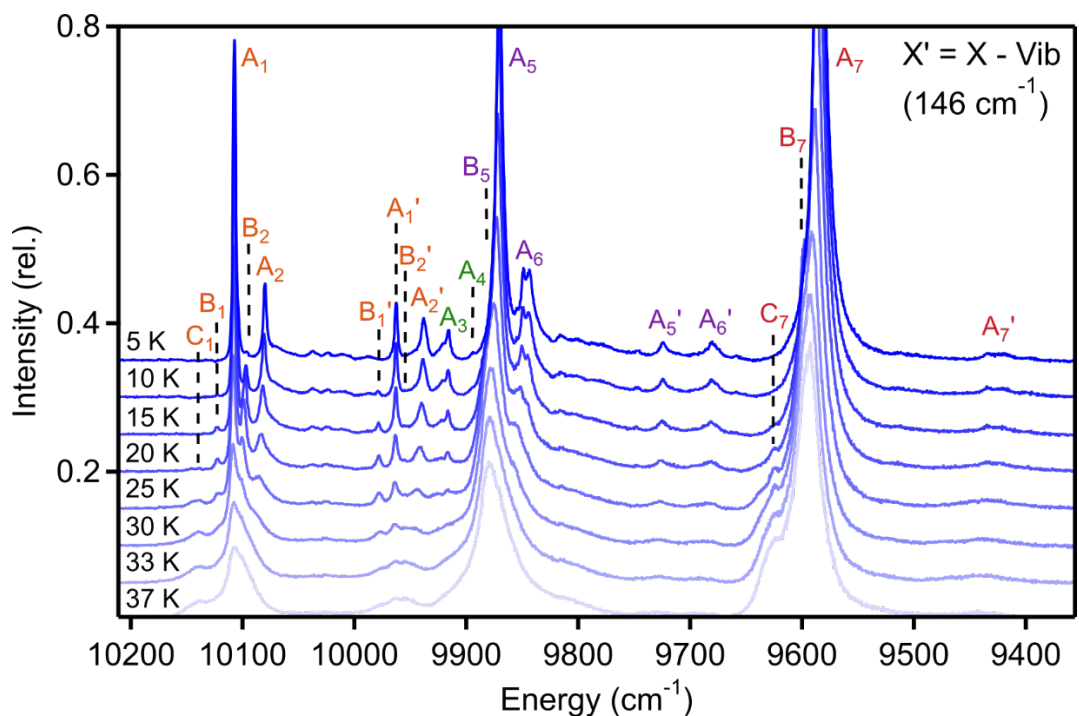

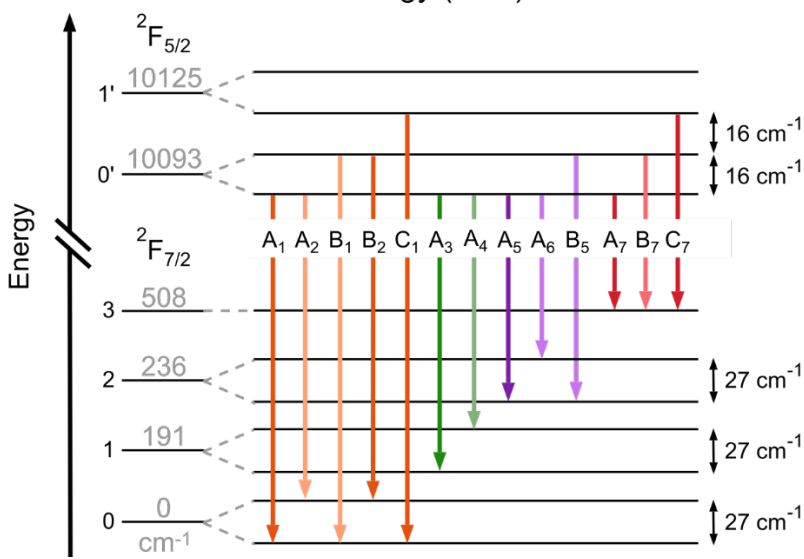

**Figure S2.** Full assignment of the CrBr$_3$:Yb$^{3+}$ PL spectrum. At 5 K, only the lowest-energy excited state (0') is populated and only peaks labeled A are observed. As the temperature increases, B peaks arise from the higher-energy spin level of the exchange-split 0' state. At even higher temperatures, C peaks appear, arising from the 1' excited state. At a fixed energy spacing of ~145 cm$^{-1}$ from the purely electronic transitions, vibronic transitions are observed. These peaks are labeled with a prime ('). Another set of vibronic peaks is possibly observed between the main and primed peaks. The 5 and 20 K spectra are plotted on a logarithmic y-axis in Figure S3 below.

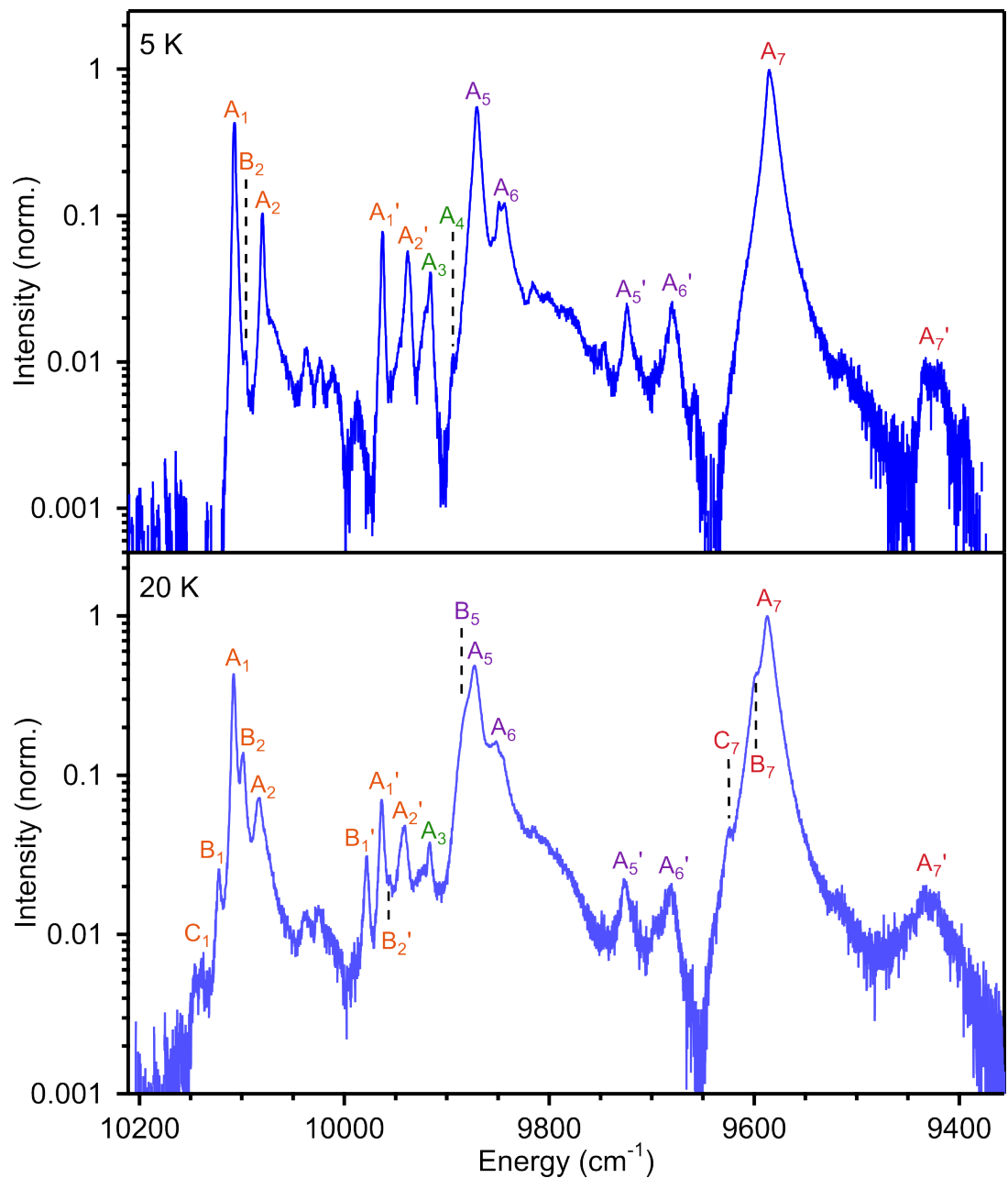

**Figure S3.** Full $CrBr_3:Yb^{3+}$ PL spectrum collected at 5 K (top) and 20 K (bottom), plotted on a logarithmic y-axis to facilitate identification of weak peaks. Peaks are labeled according to the diagram shown in Figure S2.

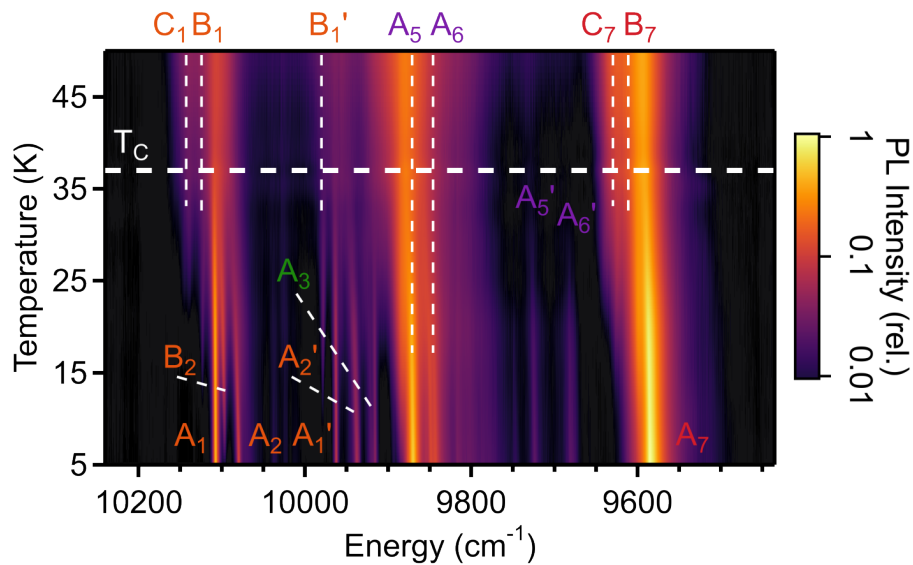

**Figure S4.** False-color plot of the CrBr$_3$:Yb$^{3+}$ 0.4 % spectrum plotted versus temperature. At higher temperatures, the center of mass of the peaks blueshifts due to the appearance of hot bands.

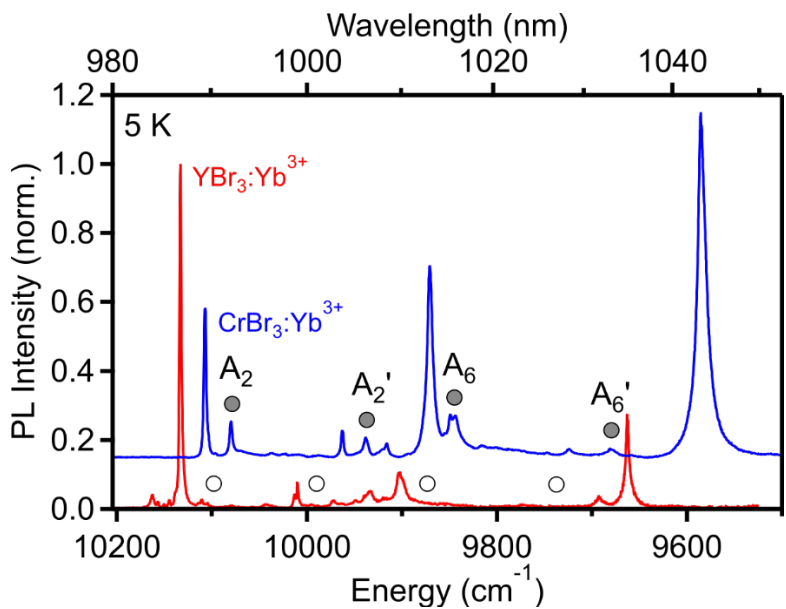

**Figure S5**. 5 K PL spectra of CrBr$_3$:Yb$^{3+}$ (blue) and YBr$_3$:Yb$^{3+}$ (red). Several peaks associated with exchange splitting are indicated by filled circles for CrBr$_3$:Yb$^{3+}$. Prime symbols (') indicate vibronic features. The absence of the corresponding peaks in the non-magnetic YBr$_3$:Yb$^{3+}$ is indicated by empty circles. The CrBr$_3$:Yb$^{3+}$ spectrum is offset vertically for easier comparison.

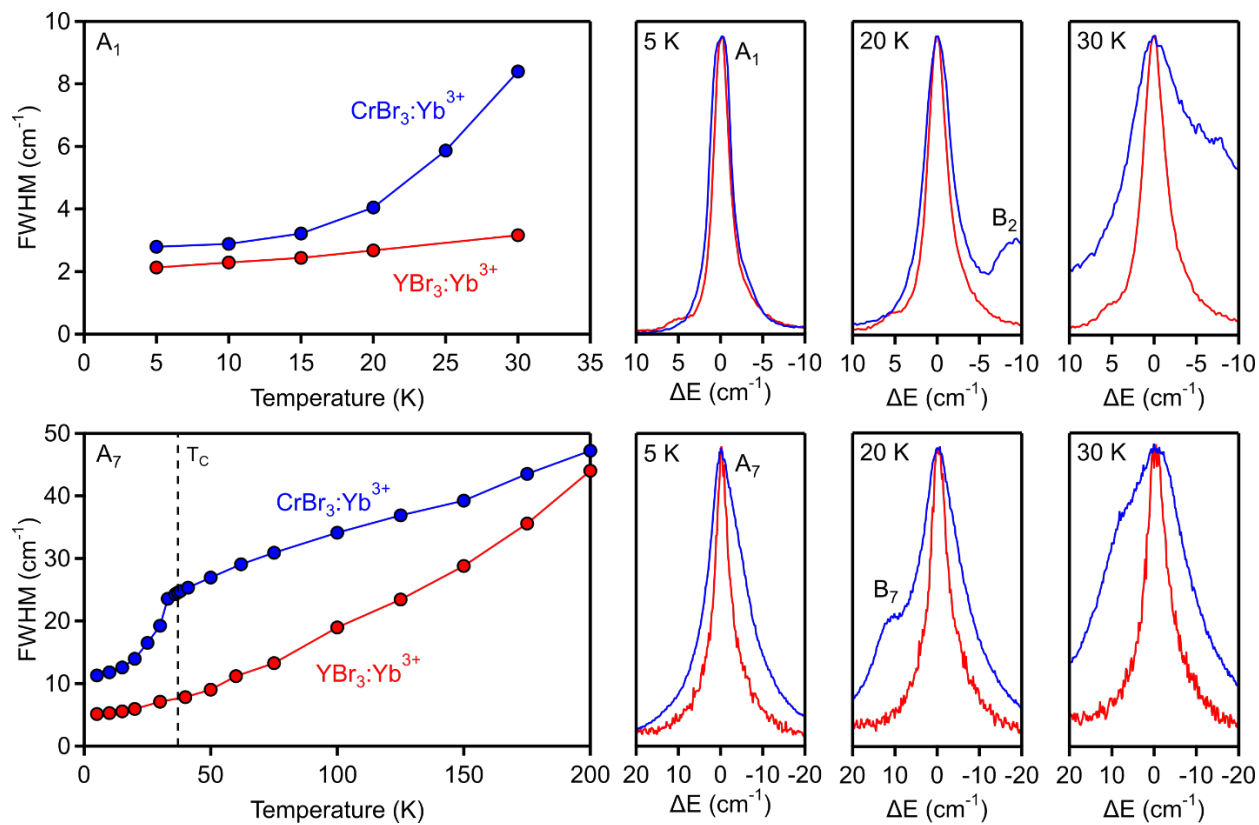

**Figure S6. Top:** Comparison of the full width at half maximum (FWHM) of the highest-energy peak ($A_1$) in both $CrBr_3$:$Yb^{3+}$ 0.4% (blue) and $YBr_3$:$Yb^{3+}$ (red) at various temperatures. The $CrBr_3$:$Yb^{3+}$ FWHM increases more rapidly than that of $YBr_3$:$Yb^{3+}$ due to spin fluctuations. Offset spectra of both are shown at select temperatures for visual comparison. Above ~30 K, the $CrBr_3$:$Yb^{3+}$ $A_1$ peak broadening is too large to quantify reliably. **Bottom:** FWHM of the lowest-energy peak ($A_7$) in both $CrBr_3$:$Yb^{3+}$ 0.4% (blue) and $YBr_3$:$Yb^{3+}$ (red) at various temperatures. A clear discontinuity is seen at $T_C$ in the $CrBr_3$:$Yb^{3+}$ FWHM that is attributed to magnetic ordering of the $Cr^{3+}$ lattice. The reduction in $Cr^{3+}$ nearest-neighbor spin fluctuations results in a sharpening of the $Yb^{3+}$ peaks at lower temperatures. The absence of peaks $B_2$ (top) and $B_7$ (bottom) in the $YBr_3$:$Yb^{3+}$ spectrum is further evidence that the $CrBr_3$:$Yb^{3+}$ spin states are split *via* exchange interactions (see Figure S2 for peak assignment). Note that unlike the $A_1$ peak, the exchange splitting of this $A_7$ peak in $CrBr_3$:$Yb^{3+}$ is not resolved and its temperature dependence is thus embedded in these linewidths.

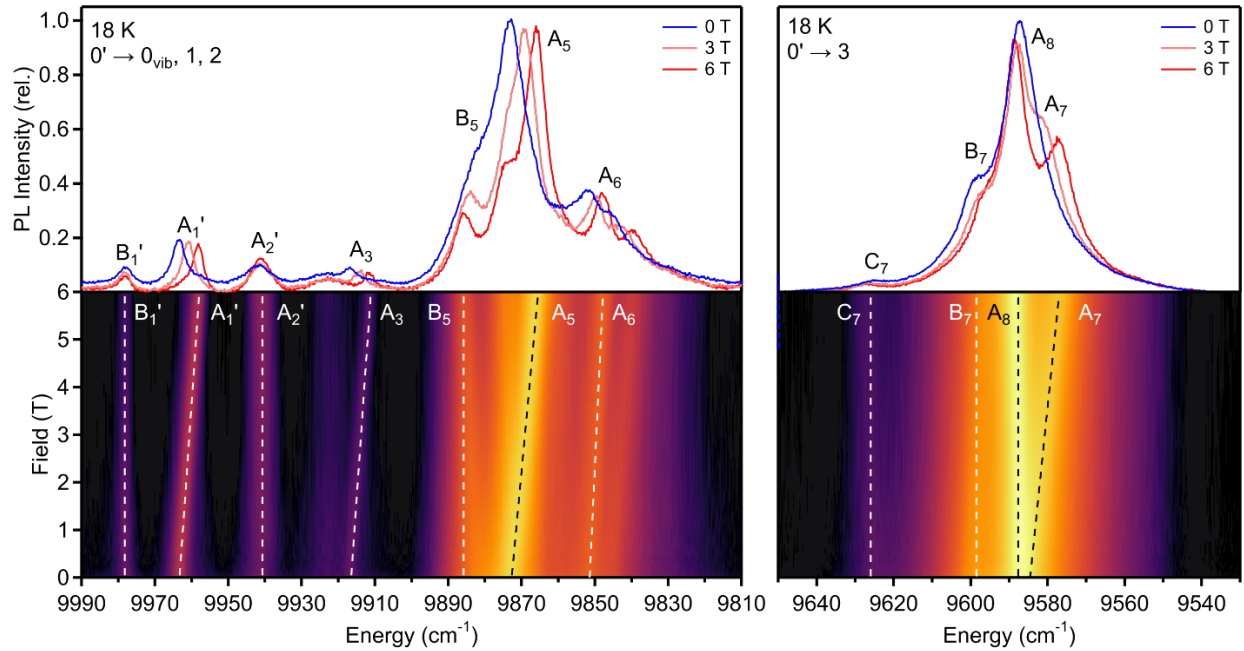

**Figure S7.** Top: PL spectra of $CrBr_3$:$Yb^{3+}$ 0.4% measured in the region of the 0' → $0_{vib}$, 1, 2, and 3 transitions at 18 K and at various external magnetic fields ($B_{ext}$ || $c$). Bottom: False-color plots of the PL in the same region vs $B_{ext}$. See Figure S2 for peak assignments. Note that peaks $A_7$ and $A_8$ are (nearly) degenerate ($\Delta E_{exch} \approx 0$ cm$^{-1}$) at $B_{ext} = 0$ T and both are labeled together as $A_7$ in other figures, where only a single peak is visible. At higher external magnetic field the degeneracy is lifted and $A_7$ shifts to lower energies like the other odd-value A peaks, while $A_8$ stays in place like the even-value A peaks.